\documentclass[11pt]{article}
\usepackage[latin1]{inputenc}  
\usepackage{latexsym}
\usepackage{amsmath,amsthm,amssymb,lscape}
\usepackage{pdflscape}
\usepackage[dvips]{graphicx, color}
\usepackage{multicol}
\usepackage{chicago}               

\usepackage[active]{srcltx}

\usepackage{url}

\makeatletter
\RequirePackage{verbatim}
\def\boxedverbatim{%
  \def\verbatim@processline{%
    {\setbox0=\hbox{\the\verbatim@line}%
    \hsize= d0 \the\verbatim@line\par}}%
  \@minipagetrue
  \@tempswatrue
  \setbox0=\vbox
  \bgroup\small\verbatim
}
\def\endboxedverbatim{%
  \endverbatim
  \unskip\setbox0=\lastbox
  \egroup
  \fbox{\box0}
}

\makeatletter

\begin{document}


\section*{Comment on `Open is not forever: a study of vanished open access journals'} 
Gerta R\"ucker\\
Institute of Medical Biometry and Statistics, Medical Center -- University of Freiburg\\
Stefan-Meier-Strasse 26\\
D-79115 Freiburg, Germany\\
e-mail: ruecker@imbi.uni-freiburg.de


\vspace{1cm}

This is a comment to an article by Laakso, Matthias and Jahn (arXiv:2008.11933).

\vspace{1cm}

With interest I read an article published by Laakso et al. \cite{laakso2020opena} that was already commented on by Shelomi \cite{shelomi2020commenta}. Laakso et al. reported a study in which they found 176 Open Access (OA) journals that had vanished from the web between 2000 and 2019. As Shelomi noted, it was unfortunate that Laakso et al. did not distinguish between predatory and reputable journals, and I agree. In my present comment, however, I want to point to another weakness of the article by Laakso et al., which is the statistical analysis. They consider their sample of 176 vanished journals and present some descriptive statistics on their characteristics, such as the time lag between the last publication and vanishing (their Figure 3 and Table 3), the distribution of academic disciplines (their Figure 4), their affiliation (whether academic or not), and the geographic distribution (Figure 5 and Table 4). Of course, this is legitimate as a description of a selected sample. However, the authors, ignoring that their sample was selected conditionally on a journal having vanished from the web, misinterpret their findings. Their statement that `our study provides valuable insight into the types of OA journals that are especially at risk of vanishing' (page 22) is not justified by evidence, simply because a risk can only be quantified on a population basis. Otherwise, any analysis can only be descriptive.

Laakso et al. found that `North America and South Asia represent a disproportionately larger share of vanished than active OA journals' and interpreted this as `journals published in North America \dots belong to the high risk group'. For the geographic distribution this is justified, as they refer to a list of active OA journals as a control group (Table 4). However, they use the term `risk' also for other factors, writing for example: `Moreover, our study provides valuable insight into the types of OA journals that are especially at risk of vanishing', then referring to journals affiliated with academic institutions or scholarly societies, without providing the numbers of active OA journals for comparison. This might provoke a well known and common error: the inversion fallacy, that is, confusing the denominators. To find out whether scholarly-based journals are at higher risk of vanishing than other journals, it is not correct to compare the proportions of journals by affiliation \emph{within vanished journals}. It would be necessary to compare vanished and active journals and to adjust for covariates in a multivariable analysis. The limitation of this study as a case series should have been acknowledged by the authors.


\end{document}